# Radially symmetric and azimuthally modulated vortex solitons supported by localized gain

Valery E. Lobanov, Yaroslav V. Kartashov, Victor A. Vysloukh, and Lluis Torner

*ICFO-Institut de Ciencies Fotoniques, and Universitat Politecnica de Catalunya, Mediterranean Technology Park, 08860 Castelldefels (Barcelona), Spain*

We discover that a spatially localized gain supports stable vortex solitons in media with cubic nonlinearity and two-photon absorption. The interplay between nonlinear losses and gain in amplifying rings results in suppression of otherwise ubiquitous azimuthal modulation instabilities of radially symmetric vortex solitons. We uncover that the topology of the gain profile imposes restrictions on the maximal possible charge of vortex solitons. Symmetry breaking occurs at high gain levels resulting in the formation of necklace vortex solitons composed of asymmetric bright spots

Vortex solitons are localized nonlinear excitations which carry screw phase dislocations and nonzero angular momentum [1]. In most of uniform media with local nonlinearity such states are prone to the azimuthal modulation instabilities. Various strategies have been developed to stabilize vortex solitons in conservative media, which include competing nonlinearities [2,3], optical lattices [4-6], and nonlocality of the nonlinear response [7]. Vortex solitons also appear in dissipative systems [8,9]. Stable dissipative vortex solitons were found in laser amplifiers [10] and in systems governed by the complex Ginzburg-Landau equation [11,12]. Optical lattices substantially enrich the properties of solitons in such settings [13].

The formation of dissipative solitons is strongly affected by a spatially modulated gain. In one-dimensional gain landscapes this phenomenon was studied in waveguides [14], optical lattices [15], and Bose-Einstein condensates [16]. In such systems solitons form in regions with locally increased gain and they can be stable because instability of zero background is suppressed [14-16]. The implications of such effect in two-dimensional geometries have never been addressed.

In this Letter we study vortex solitons supported by localized gain landscapes and two-photon absorption in media with cubic nonlinearity. We find that ring-shaped gain profiles support stable radially symmetric vortex solitons. The azimuthal gain modulation imposes restrictions on the maximal possible charge of the vortex solitons. Also, we show that suitable gain landscapes support necklace-like vortex solitons featuring asymmetric bright spots.

We consider the propagation of laser beam along the $\xi$-axis in a cubic nonlinear medium with two-photon absorption and localized gain described by the equation:

$$i\frac{\partial q}{\partial \xi} = -\frac{1}{2}\left(\frac{\partial^2 q}{\partial \eta^2} + \frac{\partial^2 q}{\partial \zeta^2}\right) - q|q|^2 + i\gamma(\eta,\zeta)q - i\alpha q|q|^2. \qquad (1)$$

Here $\eta, \zeta$ and $\xi$ are the normalized transverse and longitudinal coordinates, respectively; the function $\gamma(\eta,\zeta)$ describes the transverse gain profile; $\alpha$ is the parameter of two-photon absorption. Eq. (1) can be used to describe the nonlinear response of semiconductor materials where soliton formation occurs for wavelengths below half the band-gap and the two-photon absorption dominates [17]. Semiconductor materials are used for fabrication of optical amplifiers with high optical gain [18]. Note that ring-shaped gain profiles may be realized by using suitable pump beams or by shaping the concentration of active centers.

First we address vortex solitons supported by an amplifying ring $\gamma(r) = p_i \exp[-(r-r_c)^2/d^2]$, where $p_i$ is the gain parameter, $r = (\eta^2 + \zeta^2)^{1/2}$, $d$ and $r_c$ are the width and radius of amplifying ring, respectively. We set $r_c = 5.25$ and $d = 1.75$. We search for vortex solitons of the form $q(\eta,\zeta,\xi) = w(\eta,\zeta)\exp(im\varphi)\exp(ib\xi)$, where $b$ is the propagation constant, $\varphi$ is the azimuthal angle and $m$ is the topological charge. A first important result of this Letter is that the system described by (1) with a ring gain profile supports stable radially symmetric vortex solitons. The competition between localized gain and nonlinear losses results in simultaneous suppression of collapse and azimuthal modulation instabilities. The profiles of such radially symmetric vortices can be obtained using direct propagation governed by Eq. (1) up to $\xi \sim 2000$ with the input $q|_{\xi=0} = r^m \exp(-r^2 + im\varphi)$. Note that stable propagation over such huge distances is also an indication of the stability of the vortex solitons. We found that vortex solitons with charges at least up to $m = 6$ can be excited, although vortices with high charges require higher gain levels for their formation (this level decreases with growth of $r_c$ and $d$). Such solitons can be generated using a variety of ring-shaped input beams of very different widths with appropriate phase distribution, even if the initial phase dislocation is shifted by up to $0.2 r_c$ from the center of the amplifying ring. Under suitable conditions, inputs beams shade radiation away, experience fast reshaping upon propagation, and asymptotically approach stationary vortex solitons. The winding number of the input phase distribution determines the topological charge of the excited vortex, while its radius is always close to that of the amplifying ring. Typical shapes of stable radially symmetric vortex solitons with $m = 1,2,3$ are shown in Fig. 1.

Since in dissipative systems solitons form when an exact balance between gain - losses and between diffraction - nonlinearity is achieved simultaneously [8,9], the propagation constant (as well as all other soliton parameters) is determined by the gain and loss coefficients $p_i$ and $\alpha$. Therefore, further we set $p_i$ as a control parameter. For fixed values of $p_i$ and $\alpha$, stable vortex solitons with different charges may coexist (Fig. 1). Vortex solitons with higher topological charges are somewhat broader. The dependencies of the energy flow $U = \int\int_{-\infty}^{\infty} |q|^2 d\eta d\zeta$ and the $\xi$-projection of the angular momentum $L = \mathrm{Im}\int\int_{-\infty}^{\infty} [\eta q^* \partial q/\partial \zeta + \zeta q^* \partial q/\partial \eta] d\eta d\zeta$ on the gain parameter are shown in Fig. 2(a). Both $U$ and $L$ (for radially symmetric solitons one has $L = mU$) are monotonically growing functions of $p_i$. At fixed $p_i$ the energy flow only slightly increases with increase of topological charge. One finds that below a threshold $p_i$ value radially symmetric vortex solitons become unstable [the dependencies $U(p_i)$ and $L(p_i)$ in Fig. 2(a) terminate in corresponding points]. These results are confirmed by linear stability analysis. The threshold value of gain coefficient increases with growth of $\alpha$, at least for $\alpha > 1.5$, [Fig. 2(b)] and decreases with increase of the radius and width of the amplifying ring. Vortex solitons with higher topolog-ical charge require higher gain levels for their stabilization. The threshold gain required for the instability suppression of vortex with $m=3$ increases with $\alpha$ much faster than the threshold gain for vortex with $m=1$.

The available new phenomena become even richer if the gain landscape is azimuthally modulated. We set $\gamma = p_\mathrm{i} \sum_{k=1}^{n} \exp[-(\eta - r_\mathrm{c} \cos\varphi_k)^2/d^2 - (\zeta - r_\mathrm{c} \sin\varphi_k)^2/d^2]$, where $\varphi_k = 2\pi(k-1)/n$. Such gain landscape consists of $n$ amplifying Gaussian channels of the width $d=1.0$ arranged into a necklace structure of the radius $r_\mathrm{c} = 0.7n$. We found that this gain landscapes support stable vortex solitons which are strongly azimuthally modulated (typical shapes are depicted in Fig. 3).

A second central result of this Letter is that an azimuthal modulation of the gain landscape imposes restrictions on the maximal charge of stable vortex solitons. This means that gain landscapes with a given discrete rotational symmetry affect the topology of the dissipative vortices somehow mimicking how optical lattices determine the topology of conservative vortex solitons [6]. The comprehensive simulations that we performed, using a variety of inputs, are consistent with the conclusion that an azimuthally modulated gain landscape with $n$ amplifying channels can support vortex solitons with topological charges $m \leq \mathrm{int}[(n-1)/2]$, where the function $\mathrm{int}(...)$ stands for the integer part. Thus, landscapes with $n=3,4$ support vortex solitons with charge $m=1$, landscapes with $n=5,6$ support vortex solitons with $m=1,2$, while for $n=7,8$ one gets $m=1,2,3$, etc. The minimal number of amplifying channels that can support azimuthally modulated vortex solitons is $n=3$. For low $p_\mathrm{i}$ values the azimuthally modulated vortex solitons are extended and the light field penetrates considerably into the absorbing domain. The bright spots in such low-amplitude vortices are symmetric [see Figs. 3(a), 3(c), and 3(e)]. However, with increase of $p_\mathrm{i}$ the symmetry breaking takes place, with individual bright spots in the vortex profile becoming asymmetric. This occurs because at high $p_\mathrm{i}$ even in a single gain channel a branch of asymmetric solitons bifurcates from the branch of symmetric states making them unstable to perturbations with azimuthal index 1. In this case the maxima of bright spots shift from the centers of the amplifying rings, while the spots experience considerable reorientation with respect to the axes connecting the centers of the amplifying channels and the point $\eta, \zeta = 0$ [see Figs. 3(b), 3(d), and 3(f)]. Therefore, another important prediction of this Letter is that azimuthally modulated gain landscapes can support stable vortex solitons with nonconventional shapes composed of highly asymmetric bright fragments. The asymmetry becomes more and more pronounced with increasing gain. Notice that in dissipative systems the solitons are characterized by the internal energy flows, a property that may result in the interesting effect when due to the symmetry breaking and bright spot reorientation a nonzero angular momentum can appear at high $p_\mathrm{i}$ values even on multipole solitons. This effect occurs for even $n$ (i.e. $n=6,8$).

Not all vortex solitons whose charges are given by the above mentioned charge rule are stable. Propagating them up to $\xi \sim 2000$ we found that vortices with highest charges can be stable, while vortices with lowest charges are usually unstable. Thus, for $n=3,4$ the only existing vortex with $m=1$ can be stable. For $n=5,6$ the vortex with $m=2$ can be stable, but vortex with $m=1$ was unstable for all parameters that we considered. The typical dependencies $U(p_\mathrm{i})$ and $L(p_\mathrm{i})$ are shown in Figs. 4(a) and 4(c). The dependence $U(p_\mathrm{i})$ [as well as the dependence $U(b)$ shown in Fig. 4(b)] may be nonmonotonic, as it occurs for $n=5$, $m=2$. In this case the entire soliton family was found to be stable, while for $n=6$, $m=2$ the $U(p_\mathrm{i})$ curve stops at the point where soliton becomes unstable. Thus, in contrast to conservative systems, the branches of dissipative solitons where $dU/db<0$ (and also $dU/dp_\mathrm{i}<0$) can be stable. Such result was found to hold also in one-dimensional settings, e.g. in [15]. Interestingly, for $n=6$ the angular momentum that monotonically decreases with decrease of $p_\mathrm{i}$ may change its sign. This occurs due to the symmetry breaking, when the local energy flows inside asymmetric bright spots start contributing to the global angular momentum. Stable azimuthally modulated vortex solitons exist above the threshold gain level [Fig. 4(d)]. The minimal gain required for the existence of stable vortex soliton rapidly increases with growth of nonlinear losses.

Summarizing, we showed that radially symmetric or azimuthally modulated gain landscapes imprinted in cubic media with two-photon absorption support rich families of stable vortex solitons. With radially symmetric gain profiles the azimuthal modulation instabilities can be suppressed for solitons with topological charges at least up to $m=6$, while azimuthally modulated gain landscapes impose severe restrictions on the stability and topological charges of vortex solitons.

# References with titles

# References without titles

# Figure captions

Figure 1.  Field modulus (top) and phase distribution (bottom) for vortex solitons with (a) $m=1$, (b) $m=2$, and (c) $m=3$ at $\alpha=2$, $p_{\rm i}=3$. White circles stand for the maxima of ring with gain.

Figure 2.  (a) Energy flow and angular momentum of $m=2$ vortex soliton versus gain parameter at $\alpha=2$. (b) Minimal gain required for existence of stable vortex solitons with $m=1$ and $m=3$ versus $\alpha$.

Figure 3.  Field modulus (top) and phase distribution (bottom) for vortex solitons with (a),(b) $n=3$, $m=1$, (c),(d) $n=5$, $m=2$, and (e),(f) $n=6$, $m=2$. Panels (a),(c),(e) correspond to $p_{\rm i}=3.2$, while panels (b),(d),(f) correspond to $p_{\rm i}=5.5$. In all cases $\alpha=2$.

Figure 4.  Energy flow of $m=2$ vortex solitons versus gain parameter (a) and versus propagation constant (b) at $\alpha=2$. (c) Angular momentum of $m=2$ vortex solitons versus gain parameter at $\alpha=2$. (d) Minimal gain required for existence of stable $m=2$ vortex solitons versus $\alpha$.

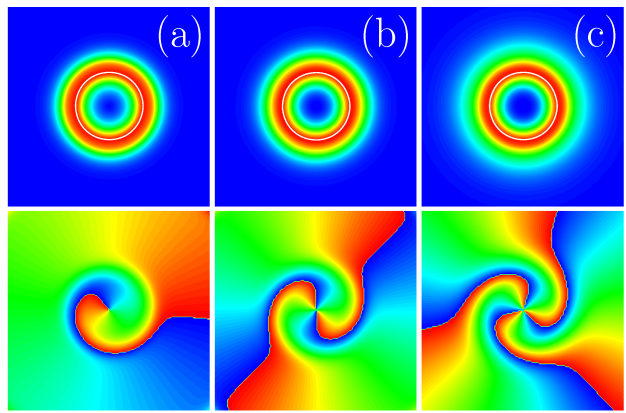

Figure 1. Field modulus (top) and phase distribution (bottom) for vortex solitons with (a) $m=1$, (b) $m=2$, and (c) $m=3$ at $\alpha=2$, $p_\mathrm{i}=3$. White circles indicate maximum of gain ring.

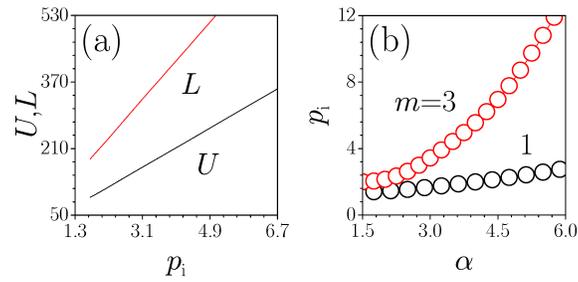

Figure 2.  (a) Energy flow and angular momentum of $m=2$ vortex soliton versus gain parameter at $\alpha=2$. (b) Minimal gain required for existence of stable vortex solitons with $m=1$ and $m=3$ versus $\alpha$.

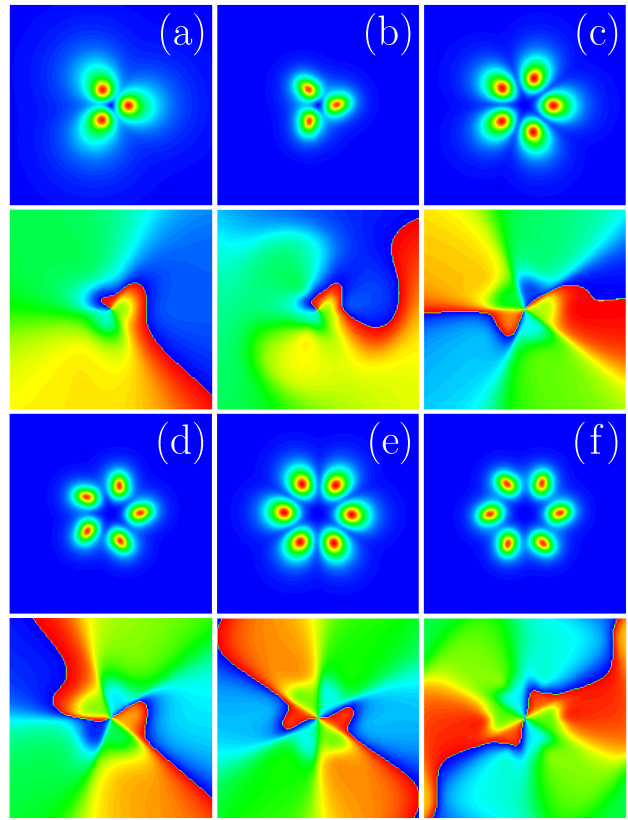

Figure 3. Field modulus (top) and phase distribution (bottom) for vortex solitons with (a),(b) $n = 3$, $m = 1$, (c),(d) $n = 5$, $m = 2$, and (e),(f) $n = 6$, $m = 2$. Panels (a),(c),(e) correspond to $p_i = 3.2$, while panels (b),(d),(f) correspond to $p_i = 5.5$. In all cases $\alpha = 2$.

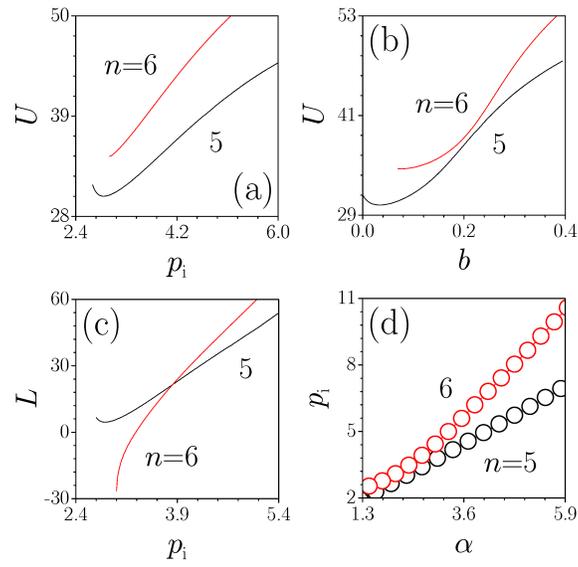

Figure 4. Energy flow of $m=2$ vortex solitons versus gain parameter (a) and versus propagation constant (b) at $\alpha=2$. (c) Angular momentum of $m=2$ vortex solitons versus gain parameter at $\alpha=2$. (d) Minimal gain required for existence of stable $m=2$ vortex solitons versus $\alpha$.

11